%% file: main.tex
\documentclass[acmsmall, manuscript,screen]{acmart}

\newcommand{\Algo}{algorithmic }
\newcommand{\Chrono}{chronological }
\newcommand{\Twitter}{Twitter/X }

\setcopyright{acmlicensed}
\copyrightyear{2024}
\acmYear{2024}
\acmDOI{~}

\acmConference[CSCW '24]{Computer Supported Cooperative Work and Social Computing}{November 9--13,
  2024}{San José, Costa Rica}
\acmISBN{978-1-4503-XXXX-X/18/06}

\begin{document}

\title[Lower Quantity, Higher Quality]{Lower Quantity, Higher Quality: Auditing News Content and User Perceptions on Twitter/X Algorithmic versus Chronological Timelines}

\author{Stephanie Wang}
\authornote{Both first authors contributed equally.}
\email{stephtw@seas.upenn.edu}
\orcid{0009-0002-3437-2227}

\author{Shengchun Huang}
\authornotemark[1]
\email{shengchun.huang@asc.upenn.edu}
\orcid{0000-0001-6581-3625}
\affiliation{%
  \institution{University of Pennsylvania}
  \city{Philadelphia}
  \state{PA}
  \country{USA}
}

\author{Alvin Zhou}
\affiliation{%
  \institution{University of Minnesota}
  \city{Minneapolis}
  \country{USA}
  }
\email{alvinyxz@umn.edu}
\orcid{0000-0001-5410-9712}

\author{Danaë Metaxa}
\affiliation{%
  \institution{University of Pennsylvania}
  \city{Philadelphia}
  \country{USA}
  }
\email{metaxa@seas.upenn.edu}
\orcid{0000-0001-9359-6090}

\renewcommand{\shortauthors}{Wang and Huang, et al.}

\begin{abstract}
Social media personalization algorithms increasingly influence the flow of civic information through society, resulting in concerns about ``filter bubbles'', ``echo chambers'', and other ways they might exacerbate ideological segregation and fan the spread of polarizing content. To address these concerns, we designed and conducted a sociotechnical audit (STA) to investigate how Twitter/X's timeline algorithm affects news curation while also tracking how user perceptions change in response. We deployed a custom-built system that, over the course of three weeks, passively tracked all tweets loaded in users' browsers in the first week, then in the second week enacted an intervention to users' \Twitter homepage to restrict their view to only the \Algo or \Chrono timeline (randomized). We flipped this condition for each user in the third week. We ran our audit in late 2023, collecting user-centered metrics (self-reported survey measures) and platform-centered metrics (views, clicks, likes) for 243 users, along with over 800,000 tweets. Using the STA framework, our results are two-fold: (1) Our \textit{algorithm audit} finds that Twitter/X's \Algo timeline resulted in a lower quantity but higher quality of news --- less ideologically congruent, less extreme, and slightly more reliable --- compared to the \Chrono timeline. (2) Our \textit{user audit} suggests that although our timeline intervention had significant effects on users' behaviors, it had little impact on their overall perceptions of the platform. Our paper discusses these findings and their broader implications in the context of algorithmic news curation, user-centric audits, and avenues for independent social science research. 

\end{abstract}

\ccsdesc[300]{Human-centered computing~Field studies}
\ccsdesc[300]{Human-centered computing~Empirical studies in collaborative and social computing}
\ccsdesc[500]{Human-centered computing~Social media}
\ccsdesc[500]{Human-centered computing~Social recommendation}

\keywords{Social media algorithms, Twitter, X, Politics, News, Sociotechnical auditing, Algorithm auditing}


\maketitle

\input{sections/1-introduction}
\input{sections/2-background}
\input{sections/3-research-questions}
\input{sections/4-method}
\input{sections/5-results}
\input{sections/6-discussion}
\input{sections/7-conclusion}

\begin{acks}
We thank our study participants for making this work possible. We thank Michelle Lam for her helpful guidance on the Intervenr system and its codebase. We thank Yphtach Lelkes, Sandra González-Bailón and CIND participants for their constructive comments and suggestions. We also thank our anonymous reviewers for their valuable feedback on our paper. Author Stephanie Wang is supported by a gift from AWS AI to Penn Engineering's ASSET Center for Trustworthy AI.
\end{acks}

\bibliographystyle{ACM-Reference-Format}
\bibliography{main}

\appendix

\input{sections/8-appendix}

\end{document}

%% file: sections/1-introduction.tex
\section{Introduction}
Social media algorithms increasingly mediate civic information for digital users~\cite{reuter_2023}. Platforms such as \Twitter\footnote{N.B.: For consistency throughout this paper and with prior research, and because its name was changed during the course of this research, we refer to the platform formerly known as Twitter, now known as X, with both names, as \Twitter.}, Facebook, and YouTube not only afford users agency in selecting content based on their preferences, but also act as algorithmic intermediaries by ranking, filtering and recommending content to users~\cite{thorson_10.1111/comt.12087}, significantly influencing their online engagement~\cite{munger_2022, guess_how_2023}. Yet despite a majority of the platform's 200+ million daily users using \Twitter as a source of news information~\cite{twitter_2022}, how \Twitter's feed ranking algorithm affects news curation remains opaque, with open questions surrounding its effects on users' behaviors and perceptions. This raises several concerns for the general public and for democracy, as algorithms may exacerbate asymmetric exposure to radical or unreliable information~\cite{youtube_radical_nyt, lazer_science_2018}, reinforce people's selectivity bias for news consumption~\cite{pariser_filter_2011}, and thereby play a role in undermining social cohesion in democratic societies~\cite{gonzalez-bailon_social_2023}.

This study aims to address a set of questions surrounding \Twitter's algorithm by comparing the effects of the \Algo timeline and the \Chrono timeline on the news content available on \Twitter, and also on users' experiences with their news information environment. By intervening in the web browsing experience of participants and enforcing a week of \Algo and \Chrono timeline displays, we audit (1) the effect that each condition has on the algorithm's outputs, in terms of news density, ideological congruence, extremity, and reliability; and (2) the effect of these different algorithmic environments on user behaviors and perceptions --- users' exposure to news, engagement with news, and attitudes and beliefs about their information environment and the platform itself.


We address these questions through a sociotechnical audit~\cite{Lam_2023} with N=243 participants, conducting a three week web-based experiment during which we collect the tweets loaded in participants' browser viewports, and make in situ interventions on their \Twitter timelines. We also deploy four surveys throughout the study to measure participants' changing perceptions over the course of the study, which was deployed in late 2023. Modeled after other sociotechnical audits~\cite{Lam_2023}, our experiment consists of a one week observational phase during which no alterations are made to users' browsing experience, and a two week intervention phase during which our system intervenes on webpage content, enforcing that each user only sees the \Algo or \Chrono timeline when visiting the \Twitter homepage for one week, then flips this condition for each participant in the following week. 



Our paper extends the sociotechnical audit method to study the effect that different types of algorithm-curated social media feeds have on (1) the \textit{news content} output of the timeline, and (2) the \textit{user}, in terms of behavioral and perceptual changes driven by different content. Over our three-week study we collect over 800,000 tweets seen by our 243 participants. We report findings from the observational phase that corroborate prior work: the \Algo timeline reduces the number of tweets with external news links, and partisans on different sides of the aisle show differences in terms of exposure to cross-cutting news. We also present experimental evidence from the intervention phase showing that the \Algo timeline exposes users to less news content, but that the sources of that content are significantly less ideologically congruent, significantly less extreme, and marginally more reliable than those shown in the \Chrono timeline. This suggests that although \Twitter's timeline curation algorithm decreases the amount of news users see, the algorithm may also have a politically moderating effect on the news content shown on the platform. Through behavioral and survey data we also find that although users are sensitive to differences in the ideological bias of the news they are exposed to (and despite our intervention making swift and sizeable changes to their \Twitter content), their self-reported perceptions and attitudes largely remain stable throughout the experiment. In other words, we identify a gap between behavior and perception in response to algorithmic changes, which suggests that attitudes are more resilient to interventions in algorithmic content. Our paper discusses this divergence in user behavior and perception in response to algorithmic change, and we conclude by suggesting opportunities for future user-centered research to evaluate not just the technical output, but also the user effects of algorithmic systems.

%% file: sections/2-background.tex
\section{Related Work}
\label{relatedwork}
Our study aligns with the existing body of research investigating the impact of algorithmic curation on individuals' online information behaviors, following scholarly concerns of ideological segregation and misinformation on social media sites~\cite{gonzalez-bailon_social_2023, lazer_science_2018}. However, this area of study has encountered two primary challenges --- an inability to experimentally manipulate algorithms for real users~\cite{bruns2019filter} and restricted access to user metrics data~\cite{davidson2023platform} --- which have limited the capacity of researchers to directly address algorithmic effects on users. Motivated by existing work in algorithm auditing and algorithmic bias, this study aims to advance methodology and examine how algorithms affect users' behaviors and perceptions in the \Twitter ecosystem. We first review the existing research on algorithm audits followed by that on personalization algorithms before articulating the research questions and hypotheses. 

\subsection{Algorithm Audits}
Algorithm audits are empirical studies that systematically investigate potential biases inherent in socio-technical systems through iterative input queries and observations of algorithmic outputs~\cite{metaxa_auditing_2021}. Such methods have been widely employed to study social media platforms such as YouTube, Facebook, \Twitter, and search engines such as Google and Bing~\cite{bandy_review_2021}. Through algorithmic audits, studies have investigated societal problems including social discrimination~\cite{lambrecht2019algorithmic, sandvig2014auditing}, radicalization~\cite{ribeiro_auditing_2021}, ideological amplification~\cite{huszar_algorithmic_2022}, and political polarization~\cite{metaxa2019search, robertson_auditing_2018} on these platforms.

\subsubsection{Sock puppet audits}
A large body of the previous work audits algorithms with \textit{sock puppet} accounts: artificial user accounts created for research purposes~\cite{sandvig2014auditing}. For instance, \citet{haroon_auditing_2023} trained sock puppets into one of the five ideologies from left to right to examine the ideological bias and radicalization in YouTube recommendations.

Auditing \Twitter, \citet{bartley_auditing_2021} examined the personalization algorithms with eight bot accounts, finding that the platform's algorithms display recency and popularity biases when curating friends' posts. Similarly, researchers have emulated archetypal users to investigate how the \Twitter timeline curates news and political content, finding that algorithmic curation deprioritized external links but slightly favored low-quality news sources~\cite{bandy_diakopolous_2021_quality, bandy_diakopoulos_2021_curation}. Moreover, they found that algorithmic curation increased source diversity but amplified topical clusters of political content. 

A major advantage of sock puppet auditing is more precise control over ``user'' behavior, a factor known to affect algorithm personalization~\cite{liu_personalized_2010}. However, this comes at the cost of ecological validity, since sock puppet findings may not replicate with real human users~\cite{bandy_review_2021}.  

\subsubsection{Audits with real users}
Recent audits have increasingly begun involving real users in data collection~\cite{huszar_algorithmic_2022, guess_how_2023, robertson_users_2023}. 
Since researchers often lack access to modify platforms' algorithms, institutional support and/or collaboration are usually essential for these studies. For example, \Twitter randomly excluded 1\% of all users as a control group whose timelines remained in chronological order, such that researchers were able to conduct a quasi-experiment, finding \Twitter amplified political content with an asymmetric effect for right-wing sources~\cite{huszar_algorithmic_2022}. In another example, \citet{guess_how_2023} collaborated with Meta to examine the algorithmic effects on users' behaviors and attitudes, randomly assigning Facebook and Instagram users to either the algorithmic or chronological feed condition. Using this powerful experimental setup at large scale, researchers found some behavioral differences between users in the algorithmic and chronological feed conditions, but did not detect attitudinal changes during the period~\cite{guess_how_2023}. 

Such collaborations with companies can draw criticism regarding researcher independence~\cite{doi:10.1126/science.adi2430}, and are also largely out of reach for most research teams. 
To overcome these issues, some research groups have developed custom software to collect users' platform activities and conduct audits~\cite{Lam_2023,robertson_users_2023,robertson_auditing_2018}. In such studies, which inform our own, researchers have invited crowdworkers to install browser extensions that automatically collect data like HTML snapshots of specific pages~\cite{robertson_users_2023} or ad content loaded in the browser~\cite{Lam_2023}. 

\subsubsection{Sociotechnical auditing}
One recent algorithm audit with real users by \citet{Lam_2023} coined the term ``sociotechnical audit'' or STA, to describe audits that combine a traditional algorithm audit's focus on understanding algorithmic content with user-focused methods like randomized controlled experiments (they term the latter ``user audits''). Motivated by a lack of prior work at the intersection of algorithmic content and its effects on users, we conduct a sociotechnical audit of \Twitter. In keeping with prior work, we recruit crowdworkers who install our custom software (as described in Section~\ref{method}), after which we both collect a baseline of their \Twitter timeline. We then conduct an intervention to causally explore the impact on algorithmic content (timeline news content) and on users (measures like news trust and engagement) comparing chronological and algorithmic feeds. 

\subsection{Online News Environments and Algorithmic Effects}

Understanding online news behaviors is a major research agenda in multiple fields, including political science~\cite{guess_almost_2021,peterson_partisan_2021,guess_how_2023,messing_selective_2014}; information science~\cite{bakshy_exposure_2015,cinelli_echo_2021}; communication~\cite{yang_exposure_2020,zhou_puzzle_2023,gonzalez-bailon_asymmetric_2023}; economics~\cite{gentzkow_ideological_2011}; and computer science~\cite{haroon_auditing_2023, robertson_users_2023}, specifically including CSCW~\cite{robertson_auditing_2018,metaxa2019search,bandy_diakopoulos_2021_curation}. The theoretical foundation of this line of study is rooted in democratic norms, and the idea that obtaining unbiased and reliable news is necessary for citizens to engage in civic life~\cite{carpini_what_1996}. Correspondingly, the rise of personalization algorithms raises several scholarly concerns regarding biases in online news ecosystems, including filter bubbles (algorithms that only show users ideologically congruent content)~\cite{pariser_filter_2011, bakshy_exposure_2015, gonzalez-bailon_asymmetric_2023}, extreme content~\cite{ribeiro_auditing_2021,haroon_auditing_2023}, and misinformation~\cite{lazer_science_2018, hussein_measuring_2020}. 
Below, we explore those three areas, as well as prior work on news engagement and user perceptions in order to situate our later findings, which investigate each.  

\subsubsection{Ideological congruence}
One of the major concerns of algorithmic curation is its potential to create filter bubbles --- silos where users see information that is aligned with their preexisting beliefs~\cite{pariser_filter_2011}. Research to date has shown mixed evidence regarding this concern, finding that people are exposed to considerable cross-cutting information online ~\cite{bakshy_exposure_2015, flaxman_filter_2016}, but that personalization algorithms do amplify like-minded content~\cite{guess_how_2023, haroon_auditing_2023}. While previous studies have shown YouTube and Facebook's algorithms favor ideologically congruent content, direct evidence is missing for \Twitter, especially in the time following the platform's takeover by Elon Musk. Indirect evidence from the previous work suggests Twitter/X algorithms tended to amplify partisan content, especially for right-wing sources~\cite{huszar_algorithmic_2022}. In a sock puppet audit, researchers also found some evidence suggesting the partisan differences would be exacerbated by Twitter/X's algorithms~\cite{bandy_diakopoulos_2021_curation}. Inspired by sock puppet results, we hypothesize that personalization algorithms will increase users' exposure to ideologically congruent news sources on \Twitter. 

\subsubsection{Extreme content}
Radicalization refers to the concern that algorithmic curation constantly exposes users to extreme or resentful information~\cite{youtube_radical_nyt}. Many studies have investigated YouTube video recommendations in response to such concerns, finding consistent radicalization results for a small proportion of politically extreme users~\cite{haroon_auditing_2023,ribeiro_auditing_2021,ledwich2019algorithmic}. When probing other platforms, researchers also found some evidence indicating algorithmic bias in terms of extreme content. For instance, researchers showed that content in the Facebook algorithmic feed condition came from less moderate sources and contained more uncivil content and slur words~\cite{guess_how_2023}. Research on \Twitter remains open; some researchers have found that the platform does not amplify far-right or -left content~\cite{huszar_algorithmic_2022} while others have found that it does amplify emotional content and outgroup animosity~\cite{milli_twitters_nodate}. Based on prior work on other platforms, as well as the public perception in the post-Musk era~\cite{musk_2022}, we hypothesize that users with the algorithmic \Twitter feed will also see more extreme news sources. 

\subsubsection{News reliability}
A third line of study is misinformation, concerning whether algorithmic curation propagates unreliable news content~\cite{hussein_measuring_2020,srba_youtube_10.1145/3568392,10.1145/3616088,fernandez_analysing_2021}. Audit studies on YouTube have documented feedback loop effects, in which users who view misinformation videos are recommended more in the future~\cite{hussein_measuring_2020,srba_youtube_10.1145/3568392}. Other work has investigated how different recommendation algorithms amplify misinformation based on \Twitter user metrics~\cite{fernandez_analysing_2021,10.1145/3616088}. However, little work on the platform takes a user-centric approach to this question; in this work, we examine whether Twitter/X's algorithmic feed amplifies real users' exposure to unreliable news. 

\subsubsection{News exposure and engagement}
Fourth, we attempt to characterize user exposure to news on \Twitter. Since social media algorithms tend to deprioritize political content and news \cite{guess_how_2023, huang_auditing_2024}, specifically tweets with external links on \Twitter \cite{twitter_punish_links}, we hypothesize that people will see less news in \Twitter's \Algo timeline condition. In addition to the concerns regarding news exposure, research interest also focuses on users' engagement with news content online. Engagement behaviors are defined as comments, likes, and reshares of social media posts~\cite{lee_advertising_2018}. \citet{guess_how_2023} found Facebook and Instagram users had significantly more news engagement on the platform under the algorithmic feed condition than in the chronological feed condition. The recent code release of Twitter/X's recommendation algorithms also suggests engagement data is heavily prioritized~\cite{twiter_algo_2023}. Given this emphasis on engagement --- a central logic of the audience attention economy --- we hypothesize that this will also affect news delivery, with Twitter/X's algorithmic feed increasing people's engagement with news on the platform. 

\subsubsection{Users' perceptions}
Our final area of focus is users' perceptions related to social media algorithms, since these perceptions may be even more important than reality in shaping their attitudes and behaviors~\cite{van_setten_support_2017}. Existing work suggests laypeople are aware of personalization algorithms and able to make sense of them from day-to-day interactions, though these perceptions may not always be factually true~\cite{eslami_i_2015, devito_algorithms_2017,rieder_ranking_2018,bandy_diakopoulos_2021_curation}. To investigate this topic in the context of \Twitter and news media, we collect content and behavioral data on users' perceptions of different timelines alongside surveys asking users about their perceptions of the system. 

%% file: sections/3-research-questions.tex
\section{Research Questions \& Hypotheses}
\label{sec:rqs}
Structuring our study as a sociotechnical audit as established in prior work, we separate our research questions and hypotheses into three key segments: first, results from our observation-only control week; second, from our two weeks of intervention, algorithm audit results (focusing on the different timelines' impacts on news content on \Twitter), and user audit results (speaking to the timelines' impacts on users' behaviors and perceptions). Having described our motivations for these questions and hypotheses above (Section \ref{relatedwork}), we enumerate them below for clarity and ease of reference. Each of these research questions and hypotheses were also pre-registered before the study.~\footnote{Anonymized preregistration is available at \texttt{https://osf.io/5gbfd?view\_only=b56b4bae74a5417d902efc95e24a060e}}

\subsubsection*{Observational Results}
The following research questions (for which we did not have explicit hypotheses) pertain to our observational data collection, and users' normal pre-intervention browsing behavior:
\begin{itemize}
    \item[\textbf{RQ1}] What fraction of tweets normally viewed come from the algorithmic and chronological timelines, respectively?
    \item[\textbf{RQ2}] How much news do users normally see, and how biased are the news sources?
\end{itemize}

\subsubsection*{Intervention Results: Algorithm Audit}
The following hypotheses pertain to our algorithm audit, which speaks to the effect of different feeds (algorithmic and chronological) on \Twitter news content:
\begin{itemize}
    \item[\textbf{H1}] (News density) The algorithmic timeline has a \textit{lower density of news} content relative to chronological.
    \item[\textbf{H2}] (Ideological congruence) Algorithmic curation \textit{amplifies ideologically congruent news} content relative to chronological.
    \item[\textbf{H3}] (News extremity) Algorithmic curation \textit{amplifies extreme news} content relative to chronological.
    \item[\textbf{H4}] (News reliability) Algorithmic curation \textit{amplifies unreliable news} content relative to chronological.
\end{itemize}

\subsubsection*{Intervention Results: User Audit}
Finally, the remaining hypotheses and questions speak to differences in users' experiences driven by the different feeds, including behavioral differences (for which we had explicit hypotheses) and perceptual ones (for which we did not): 
\begin{itemize}
    \item[\textbf{H5}] (User exposure) Users will see \textit{less news content} on the algorithmic timeline compared to chronological.
    \item[\textbf{H6}] (User engagement) Users will \textit{engage more} with news content on the algorithmic timeline compared to chronological.
    \item[\textbf{RQ3}] (User perceptions) How do different timelines affect users' perceptions of the platform's algorithmic curation and news ecosystem?
   
\end{itemize}

%% file: sections/4-method.tex
\section{Method}
\label{method}
Our method includes infrastructure for our sociotechnical audit, study design, participant recruitment, and surveys. 
\subsection{Sociotechnical Audit and Infrastructure}
To investigate our research questions, we followed the sociotechnical audit (STA) method as defined in prior work~\cite{Lam_2023}. 
Sociotechnical audits combine a standard algorithm audit with a user audit, to understand algorithmic content alongside the effects of that content on users. In our STA, we collected the web browsing content of consenting, compensated participants and exposed them to varied algorithmic content on the \Twitter home timeline to observe the impact on their behaviors and perceptions. This structure, and the infrastructure we describe in the sections that follow, allow us to collect fine-grained information about the content users encounter in naturalistic settings, and to enact in situ interventions on the \Twitter timeline without internal access to the platform, over an extended period of time. 

\subsubsection{Infrastructure}
To execute our STA, we developed browser-based infrastructure that allowed us to collect participants' web browsing behavior, deploy custom surveys and conduct experiments on their \Twitter timeline. The system is composed of a web application, a Chrome browser extension and a data analysis pipeline.  

The web application was developed in Python as a Django web app hosted on Heroku, a cloud application platform. Data was stored in an Amazon RDS Postgres database. The web app serves as the server and database from which we managed the experiment details and deployed logic to enact interventions for participants at the right time. The web app also has a participant-facing interface with which participants can provide consent, onboard, take surveys and view key information about their enrollment in the study throughout the study period.
  
The custom Google Chrome browser extension captures tweets that load into the participant's viewport as well as URLs embedded in the tweet and participants' engagement (whether they liked, commented on, or retweeted). The browser extension enacts the logic to alter \Twitter webpage code and enforce the \Chrono or \Algo timeline when users view their \Twitter homepage. Data is sent to and received from the web app described above. 

The data analysis pipeline consists of a Python script hosted on an Amazon EC2 instance. Because \Twitter shortens all external URLs appearing in tweets to a \texttt{t.co} domain, we configured our script to run periodically and resolve all collected URLs to their final domain, before annotating each resolved domain with bias and reliability scores published by Ad Fontes Media, a commonly-used dataset in such studies~\cite{ad_fontes_media_methodology_2023}.

\subsection{Study Design}
Next we provide details of our study design, including the study phases and participant recruitment, management, and compensation. The study, as described below, was approved by both the University of Pennsylvania and Stanford University Institutional Review Boards.

\subsubsection{Study Phases}
\begin{figure}[ht]
  \centering
  \includegraphics[width=\linewidth]{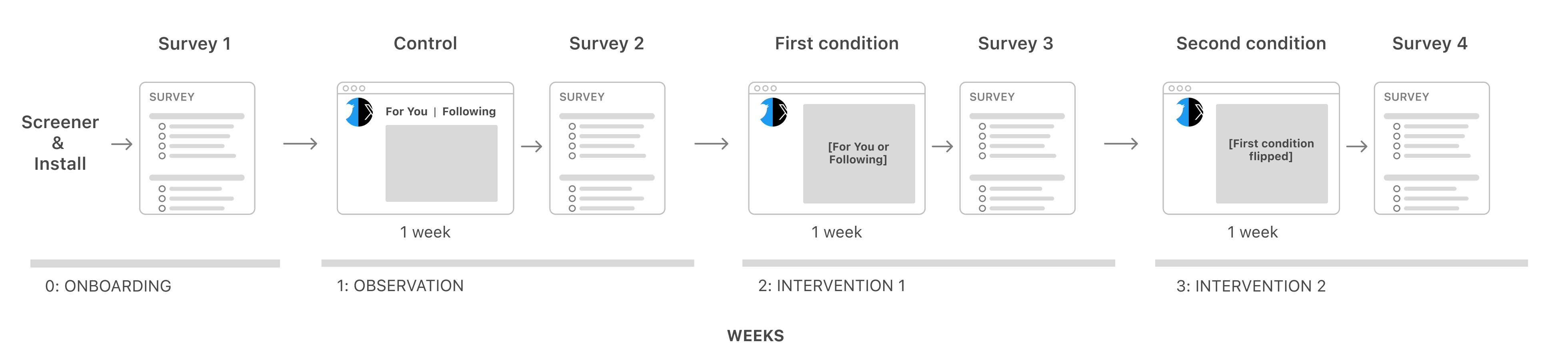}
  \Description[short]{long}
  \caption{Full timeline of the study. In the observational phase users view Twitter/X timelines as usual. In the first intervention we randomly assign users into the \Algo or \Chrono condition. In the second intervention, for each user, we flip their condition. The figure also shows the timing of the survey deployments.}
  \label{fig:timeline}
\end{figure}

Our study was designed in three phases, each lasting one week: observation, a first intervention, and a second intervention, as visualized in Figure~\ref{fig:timeline}. During the observation week we made no alterations to the user experience on \Twitter (or elsewhere) and passively collected all tweets that scrolled into view in participants' browser viewports.
We also collected views, likes, retweets, and comments on --- as well as external URLs within --- all tweets.  

\Twitter currently offers users two different timelines: a reverse-chronological one comprised only of tweets from those the user follows, called ``Following'', and an algorithmically-curated one that also includes content from others the user does not follow, called ``For You''. During the first intervention phase, we randomly assigned each participant to one of two conditions: \Algo (the ``For You'' timeline) and \Chrono (the ``Following'' timeline). Our software enforced that participants in the \Algo condition could only see the algorithmic timeline when visiting their \Twitter homepage (\texttt{twitter.com}), while participants in the \Chrono condition could only view the chronological one. In the second intervention phase we switched the randomly-assigned condition for each participant. 

\subsection{Participants}
We recruited participants from Prolific once a week during November and December 2023. A total of 822 participants onboarded and completed the background survey; of those, 243 participants were promoted to and completed the first intervention phase and therefore have their data included in our analyses. Participants from whom we did not collect any tweets during the observation phase were not promoted in the study beyond the first week observation phase. A total of 218 participants reached the end of the second intervention phase and completed all accompanying surveys. 
For our analysis we consider all participants who provided at least some data during the observational and intervention periods thus arriving at a final total of N=243 \Twitter users. Our sample was 68\% male, 62.5\% lean Democrat, with 85\% reporting some level of college education. A visual breakdown of our users' demographics can be found in Appendix~\ref{appendix:demographics}.


\subsubsection{Participant Recruitment and Management}
We first recruited participants on the Prolific academic crowdsourcing platform to take a 1-minute pre-screening survey to identify participants eligible for the main study. To be eligible, we asked that users live in the United States, be at least 18 years old, use Google Chrome as a main web browser, use \Twitter, and access \Twitter from their main web browser at least a few times per week. All participants who completed the pre-screen survey were compensated \$0.25 (approximately \$15/hour), regardless of eligibility for the main study. Eligible participants were then directed to another page notifying them of the main study hosted on our own website (\url{https://intervenr.stanford.edu/}). 

When onboarding to the main study, participants were presented with a consent page asking whether they were interested in participating in a research study that would entail sharing their browsing behavior. Consenting participants then completed a first survey in which they self-reported their partisan identification on 7-point scales consistent with ANES measures (`strong Democrat' to `strong Republican' as well as `extremely liberal' to `extremely conservative') and other demographic information including age, race, gender, household income, and level of education.

During the full three-week study, we compensated participants \$5 for each milestone requiring their active effort in the study. These were the four surveys, as visualized in Figure~\ref{fig:timeline}: onboarding (survey 1), a survey after the observation week (survey 2), a survey after the first intervention week (survey 3), and a final survey (survey 4). Each survey required about 10-15 minutes to complete, so this amounted to an estimated rate of \$20/hr. Participants received compensation via payment on Prolific. We also used Prolific's direct messaging system to communicate with participants during the study and notify them of survey tasks and offboarding procedures. We also assisted participants experiencing technical issues throughout the study, and notified participants when 72 hours elapsed without any data collected to ensure correct study enrollment.

\subsection{Survey Design}
Several of our research questions and hypotheses pertain to users' perceptions. In addition to the technical infrastructure we developed to capture behavioral methods, described above, we surveyed participants to measure these. We surveyed participants a cumulative three times (surveys 2, 3, and 4) (see Figure~\ref{fig:timeline}). In each survey we asked questions about user perceptions in the following main categories, briefly described below. For the full survey, see Appendix \ref{appendix:survey}.

\paragraph{Perceived echo chambers} While there has been extensive research on echo chambers, the literature on \textit{perceived} echo chambers is still sparse. Therefore, the following conceptualization of perceived echo chambers is in line with the measures of echo chambers from observational studies: We measured participants' perceptions of echo chambers by asking them to roughly estimate the ideological bias of the news on \Twitter in the past week for their own timeline, for a typical Democrat's timeline, for a typical Republican's timeline, and overall on \Twitter. All were answered on a 100-point feeling thermometer ranging from ``extremely liberal'' to ``extremely conservative''.

\paragraph{Media satisfaction} We measured users' satisfaction with news content by asking whether they were satisfied with the news information they saw on \Twitter in the last week, and with their experience on the platform overall, both on 5-point Likert scales (``extremely unsatisfied'' to ``extremely satisfied''). 

\paragraph{Perceived news credibility and media trust} We asked participants to estimate the credibility of the news information they see on \Twitter (from ``extremely low credibility'' to ``extremely high credibility'') and their trust of that information (from ``extremely distrust'' to ``extremely trust''), both measured on 5-point Likert scales.

\paragraph{Perception of the platform} 
To measure their perceptions about the platform, we asked participants whether \Twitter is generally neutral, or whether it instead tends to support liberals or conservatives, answered on a 5-point Likert scale (''primarily supports liberals' posts'' to ''primarily supports conservatives' posts''). We also asked about the extent of their support for the regulation of social media companies like \Twitter on a 5-point Likert scale (''strongly agree'' to ''strongly disagree'').



\subsection{Final Dataset}
Over the course of the study period we collected a total of 952,787 tweets, which resulted in 846,494 tweets from our final pool of participants after accounting for attrition; we conduct all further analyses with this dataset. Of those, 313,517 tweets were collected in the observational phase; 253,471 were collected in the first intervention phase; and 279,506 were collected in the second intervention phase. Users engaged (liked, commented, retweeted) with 1.89\% (16,014) of all tweets collected. A majority of our tweet collection (70.5\%, 596,763 tweets) were unique as determined by tweet ID, and 1.89\% (14,817) tweets contained external news URLs identified using media-bias and reliability labels provided by Ad Fontes Media~\cite{ad_fontes_media_methodology_2023}. Ad Fontes Media evaluates the U.S. media ecosystem and produces a proprietary media bias product that rates 2,796 information services' ideological slants (ranging from the most liberal at -42 to the most conservative at +42) and news credibility (ranging from the least credible at 0 to the most credible at 64). The rating has been used and validated in various HCI, communication, and general-interest publications~\cite{zhou_puzzle_2023,huszar_algorithmic_2022,haq_2022_chi}.

Our data includes tweets collected on the \Twitter timeline, as well as on other \Twitter surfaces (e.g. Profile, Explore feed, Communities), in order to fully capture the impact of our interventions. For instance, enforcing that a user see a feed that shows tweets from users they do not follow (as we did in the \Algo condition) may prompt them to view some new users' profiles and consume many more tweets in the process. This whole consumption pattern is inextricable from the intervention, so we consider all tweets seen during the study rather than restricting to timeline-only content.

\subsection{Ethics}

We took care throughout the deployment of our tool, as well as in the design and roll-out of our study, to prioritize participants' privacy and agency during their participation. This included a webpage for participants to view and redact data they did not want us to have, and fair payment (compensating over \$15 per hour by estimates of the time users would spend enrolling, onboarding, answering surveys, and offboarding). By the same motivation, although we understand the value in open science, in favor of preserving user privacy we will not be releasing any data collected from participants, since our data document their entire \Twitter timelines as well as their browsing behavior on the platform. Finally, although we are considering open-sourcing the software for other researchers to use, we are still in the process of analyzing possible risks of doing so, and thus have not included it with this paper. 

%% file: sections/5-results.tex
\section{Results}
\label{sec:results}
We report our study results next, beginning first with initial results from the observation-only week in Section \ref{results:control}, and then describing our algorithm auditing results and user audit results. Each subsection that follows corresponds to a research question or hypothesis with the same name, as laid out in Section~\ref{sec:rqs}.

\subsection{Observation Week Results}
\label{results:control}
We first characterize at a high level users' general information behaviors on \Twitter using results from the week-long observational phase of our study. On average, users saw 1,306 tweets during the observational phase; 24.0\% of their tweets were loaded in the \Algo timeline, 9.4\% in the \Chrono timeline and 66.6\% in other, non-timeline surfaces on \Twitter. Somewhat surprisingly, although users tended to get their timeline content in the engagement-optimized \Algo feed, which is also the default when visiting the \Twitter website, their behavior reflected a tendency by many users to switch back and forth between the two: 43\% of participants saw tweets from both timelines during the observational phase. Also notable was the high proportion of tweets users saw outside of their \Twitter timelines, which contrasts with the predominant focus of most prior work studying only the timeline \cite{bandy_diakopoulos_2021_curation, bartley_auditing_2021, milli_twitters_nodate}. 

In terms of differences in exposure to news domains, 1.6\% of tweets in the average \Algo timeline contained a news URL, 7.5\% in the average \Chrono timeline and 1.0\% in non-timeline contexts. This is consistent with prior work showing that \Twitter's \Algo timeline deprioritizes external links~\cite{bandy_diakopoulos_2021_curation}, and is also supported by the recent open-sourcing of \Twitter's algorithm which suggests that tweets containing external links are filtered and deprioritized in the \Algo ``for you'' timeline~\cite{twitter_ooncompetitorURLfilter_2023}. 

Examining differences in partisans' cross-cutting news exposure: in the \Chrono timeline, right-leaning users (those answering ``extremely'' to ``slightly'' conservative) saw a moderate amount of news links from left-leaning news domains (13.5\% of all news links). Meanwhile, left-leaning users saw very little news from right-leaning domains (1.5\%). We saw similar patterns in users' \Algo timeline content; right-leaning users saw 11.8\% of their news links from left-leaning domains while only 2.7\% of news links for left-leaning users were from right-leaning news domains. This asymmetry is consistent with older work on other platforms suggesting that right-leaning users are more exposed to ideologically cross-cutting content than left-leaning users~\cite{bakshy_exposure_2015}. 

\subsection{Algorithm Audit Results}
\label{results:algoaudit}

Moving on to our two weeks of intervention data and accompanying surveys, we first present algorithm auditing results that compare the \Algo and \Chrono feeds in terms of news-related metrics, including the amount of news to which users were exposed, the partisan bias of that content, and its reliability. We statistically test each of these outcome measures using mixed-effects models, in which we include participant ID as the random effect to control for differences between participants. All following analyses use user as the unit of analysis, so that each user has three observations in the dataset, corresponding to their behavioral and perceptual indicators across the three weeks.

\begin{figure}[ht]
  \centering
  \includegraphics[width=\linewidth]{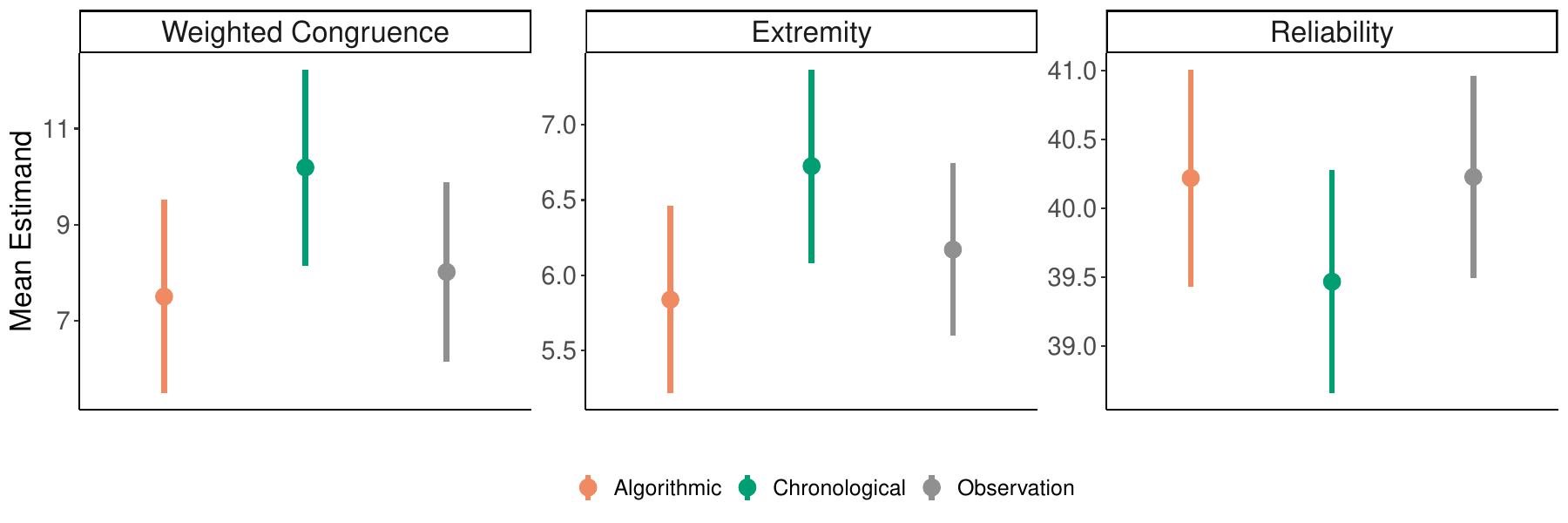}
  \Description[short]{long}
  \caption{Algorithmic effects on news curation: Users assigned to the \Algo timeline were exposed to tweets with less congruent and less extreme news links. The effect on reliability is not significant.}
  \label{fig:AlgoAuditResult}
\end{figure}

\subsubsection*{H1. News density} 
Normalizing by the number of tweets seen in each condition, we confirm our hypothesis that the \Chrono condition has a greater density of news content (3.26\%) than the \Algo condition (0.94\%).

\subsubsection*{H2. Ideological congruence} 
To measure user exposure to like-minded content, we compute the \textit{weighted congruence} score for each user's exposure to news content by multiplying the news domain's media bias and the user's ideological leaning. For example, when an ``extremely liberal'' user (ideology labeled as -3) is exposed to a news link from the New York Times (media bias labeled as -8.13 by Ad Fontes Media), we quantify the level of congruence as $(-3)*(-8.13)=24.39$. Similarly, if a ``slightly conservative'' user (ideology labeled as +1) is exposed to the same link, such instance will be quantified as $(+1)*(-8.13)=-8.13$ congruence. We find that participants were exposed to significantly more ideologically congruent news content in the \Chrono condition than in the \Algo ($\beta=2.82$, $SE=1.11$, $p=0.031$). In other words, contrary to our hypothesis, we find that the \Algo condition causes users to see less ideologically congruent news sources (Figure \ref{fig:AlgoAuditResult}).

\subsubsection*{H3. News extremity} 
To evaluate the extremity of news sources, we compute the \textit{mean absolute bias}. For example, the bias of a news link from the left-leaning New York Times, which Ad Fontes labels at -8.13 will be converted to 8.13, and a link from the far-right news site NewsMax (media bias labeled as +21.17 by Ad Fontes Media) will be kept as 21.17. We average these absolute bias scores for each user's three weeks and analyze how they have changed across the three weeks. Contrary to our hypothesis, we find that participants are exposed to more extreme news sources when in the \Chrono condition than in the \Algo condition ($\beta=1.219$, $SE=0.364$, $p=0.003$) (Figure \ref{fig:AlgoAuditResult}).

\subsubsection*{H4. News reliability}
Consistent with prior work on conservatives' news environments, our data also showed that conservatives see significantly more unreliable news at the user level, regardless of experimental condition ($\beta=-1.02$, $SE=0.15$, $p<0.001$). We had hypothesized that the \Algo feed would show less reliable news content; however, comparing the different timelines, we did not see statistically significant differences of news reliability between the two experimental conditions (Figure \ref{fig:AlgoAuditResult}).

\subsection{User Audit Results}
\label{results:useraudit}

To better understand the impact of the different feeds on users, we conducted our study as a sociotechnical audit, that pairs a user audit with a more traditional algorithm audit. Next, we outline the findings of our user audit, comparing the impact of algorithmic and chronological timelines on users' engagement with news and their perceptions of the platform from various survey measures. 

\begin{figure}[ht]
  \centering
  \includegraphics[width=\linewidth]{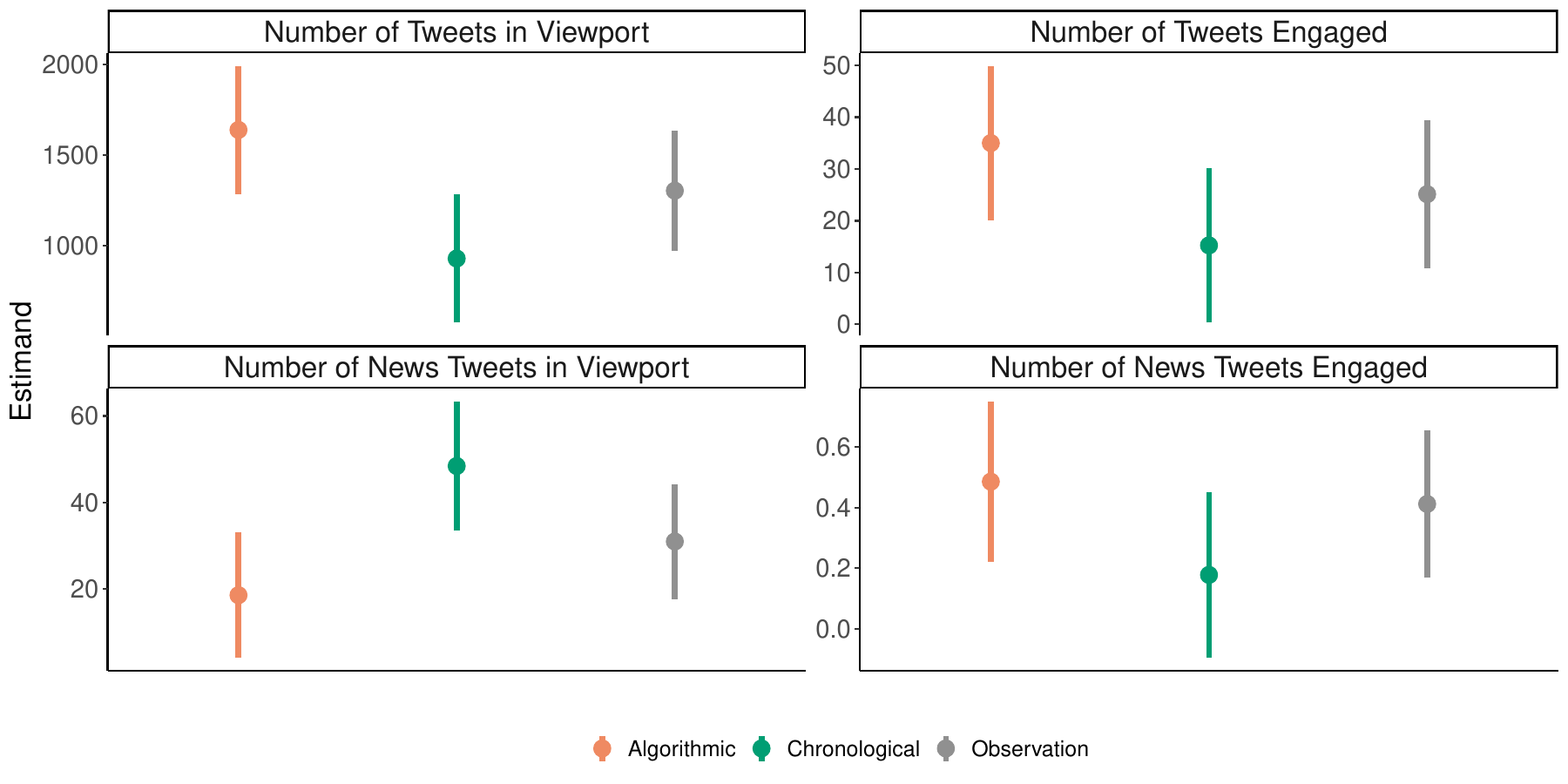}
  \Description[short]{long}
  \caption{Algorithmic effects on user behavior. Users assigned to the \Algo timeline used and engaged with \Twitter more, however, they saw less tweets with news links. The effect on their engagement with news tweets is not significant.}
  \label{fig:UserAuditResult}
\end{figure}

\subsubsection*{H5. User exposure}
We used mixed effect models to analyze our data, including participant ID as the random effect to control for differences between participants. Analyzing for the number of tweets seen by participants between the two experimental conditions, we found that the number of tweets containing news URLs was significantly higher in the \Chrono condition compared to the \Algo condition ($\beta=30.3$, $SE=8.72$, $p=0.002$). In other words, we confirm our hypothesis the \Algo timeline seems to de-prioritize showing tweets with news links to our participants (Figure \ref{fig:UserAuditResult}).

\subsubsection*{H6. User engagement}
Results showed that the number of tweets participants loaded in their browsers was significantly higher when participants were in the \Algo condition, compared to the \Chrono condition ($\beta=709$, $SE=157$, $p<0.001$). Similarly, users also engaged (i.e., liked, commented, and retweeted) more with tweets when in the \Algo condition, compared to the \Chrono condition ($\beta=19.75$, $SE=5.32$, $p<0.001$). In other words, confirming our hypothesis, users use \Twitter and engage with tweets more when they are in the \Algo condition (Figure~\ref{fig:UserAuditResult}).

Turning to engagement with tweets containing news links, however, there was no significant main effect of the different conditions on participants' like, comment, or retweet behaviors ($\beta=-0.31$, $SE=0.16$, n.s.). In other words, while the \Algo timeline led to greater user engagement overall (and fewer news links, as we saw above), user engagement with these news tweets did not significantly differ between the two conditions (Figure \ref{fig:UserAuditResult}).


\subsubsection*{RQ3. User perceptions} We report findings on user perceptions: their reported satisfaction, perceptions  of the platform, and perceptions of other users' experiences.
\paragraph{User satisfaction.} 
We conducted exploratory analyses (without explicit hypotheses) to test the effect of our timeline intervention on Twitter users' perceptions and attitudes toward the platform and media content. A repeated-measure ANOVA showed that the overall satisfaction of user experience on Twitter differed among the three conditions (observation, \Chrono, and \Algo) ($F(2, 432) = 4.54$, $p=0.011$). Users were significantly more satisfied with their user experiences during the observation (i.e., no treatment) stage ($M_O=3.38$), compared to both \Chrono ($M_C=3.23$) or \Algo ($M_A=3.21$), which did not significantly differ between each other.

\textit{User perceptions of \Twitter. }
To test whether participants who see tweets with more biased news links adjust their perceptions of the bias of their own timelines, we employed a within-subject mixed-effects model, where we regressed their perceptions (i.e., the self-reported estimate of ideological bias of news on one's own timeline) on users' behaviors (i.e., the actual mean ideological bias of news they saw on their timeline). We again include participant ID as a random effect. We found that there was a significant association between the perceptions of bias and actual news bias on Twitter/X ($\beta=0.85$, $SE=0.18$, $p<0.001$). User exposure to biased news links, however, did not change their perceptions of the average bias of all news links available on Twitter ($\beta=-0.03$, $SE=0.16$, n.s.). Such exposure also did not change their perceptions of whether the \Twitter algorithm favors conservatives or liberals ($\beta=-0.01$, $SE=0.01$, n.s.). 

We also investigated users' assessments of news credibility and media trust in the news content they see on \Twitter. With participant ID as a random effect, we regressed participants' self-reported news credibility and media trust across all three conditions on the mean reliability of the news links to which they were exposed. We found a significant and negative association, indicating that participants who saw more reliable news actually perceived \textit{lower} news credibility ($\beta=-0.02$, $SE=0.01$, $p=0.012$) and reported \textit{less} trust ($\beta=-0.03$, $SE=0.01$, $p=0.002$) in the news they saw. The conditions themselves, however, did not significantly change self-reported media trust ($F(2, 388) = 0.67$, n.s.) and user perceptions of news credibility ($F(2, 388) = 1.37$, n.s.).

\textit{Users' perceptions of other users. } 
Next, we test how users' exposure to news links changes their perceptions of how biased an average Democrat's or Republican's \Twitter timeline is. We regressed such perceptions on the mean ideological bias of the news links users were personally exposed to, controlling their own partisanship affiliation and with participant ID as a random effect. We found a significant negative main effect of the ideological bias of news exposure on their estimation of news bias on Republican timelines ($\beta=-0.51$, $SE=0.16$, $p=0.001$). In other words, exposure to more conservative news links is associated with the perception that a typical Republican's timeline was less conservative than participants previously thought. On the other hand, we found a significant and positive main effect of the ideological bias of news exposure on their estimation of news bias on Democrat timelines ($\beta=0.46$, $SE=0.20$, $p=0.017$). Put differently, exposure to more liberal news links is associated with the perception that a typical Democrat's timeline was more liberal than participants previously thought. Our association tests, by connecting news exposure during the week with news perceptions at the end of the week, cannot constitute true causal claims. However, these divergent patterns do implicate such causal possibilities in the \Twitter news ecosystem.

%% file: sections/6-discussion.tex
\section{Discussion}

The study we present here provided a sociotechnical audit of \Twitter timelines. By using a browser extension to collect real users' data and intervene in the display order of their timeline tweets (showing them one week of their \Algo and \Chrono timelines each, in random sequence), we offer the first systematic and up-to-date exploration of users' experiences using \Twitter in a post-Musk age. Notably, this study follows the recent development of methods for obtaining ecologically valid exposure data, i.e., measuring the content that real users are exposed to during natural platform use~\cite{robertson_users_2023, guess_how_2023}. Based on our observational data, we reported \textit{where} and \textit{what} people saw on \Twitter --- on average, the sampled users saw more than a thousand tweets during the observation week, with more than half of those tweets coming from non-timeline surfaces. By pairing users' self-reported ideology with the bias of news sources they were exposed to, we were able to corroborate past findings of partisan differences in news exposure (Section \ref{results:control}).
Such observational results, we believe, are valuable supplementary observations to prior work on \Twitter's information ecosystem~\cite{pfeffer_just_2023}. Also of note, given platforms' increasingly tight control over their data, including placing restrictions on data collection~\cite{davidson2023platform,freelon_computational_2018}, this study contributes valuable post-API methodology for researchers seeking to conduct external experiments on proprietary social media algorithms. 

Besides these overall and observational notes, our study also revealed significant findings pertaining to both algorithmic curation and user-level effects, which we discuss below. 

\subsection{Algorithm Audit Results}
Contrary to theoretical concerns and sock puppet audit results, the evidence we present suggests that the \Algo timeline actually played a moderating role in shaping the news content users encountered on \Twitter. Specifically, the news sources users were exposed to in the \Algo timeline condition showed characteristics of being less ideologically congruent to the user, less extreme, and slightly more reliable compared to the chronological timeline condition. This suggests that the users' ideological congruence with the news shared by those they follow is greater compared with the news prioritized by \Twitter's algorithm. Considering the institutional changes in \Twitter in the past two years, it is unclear to what extent these findings are due to intentional platform-level algorithm adjustment, and to what extent these algorithms will remain stable in the coming months and years. However, in in the event that they do remain stable, these results optimistically suggest that the algorithmic biases observed by prior work on the system may not translate into tangible real-world consequences. This point emphasizes the importance of our sociotechnical, ecologically valid study design, and the importance of considering the interconnected dynamics of human-algorithm interactions. 

Also of note, our findings diverge from those reported for Facebook, where recent prior work (also with real users) has found the algorithm does not play a moderating effect on users' news exposure~\cite{guess_how_2023}. As we observe in comparing these two studies, recommendation algorithms may operate dissimilarly across different platforms, even when they serve comparable functions. This divergence serves as a reminder for researchers to exercise caution when generalizing algorithm audit results to other online platforms.

\subsection{User Audit Results}
In our study, we also collected user-centric metrics regarding participants' experiences across timeline settings. Behavioral evidence revealed that users tended to see more tweets but less news in the \Algo timeline condition, suggesting this timeline is better able to hold users' attention but less likely to show them news content. This finding is partially consistent with the recent open-sourcing of \Twitter's recommendation algorithm~\cite{twiter_algo_2023} which explicitly claims to optimize for user engagement with algorithm-curated content. Our other major finding on user perceptions pertained to perceptions concerning both the platform and content amidst the real differences in exposure caused by our experimental interventions. Here we found that although users' perceptions of the content bias were aligned with the changes in their timelines, overall attitudes towards broader concepts such as platform bias and news trust remained the same. 

The implications drawn from our user audit results are threefold. The first point is methodological. This study, unlike most prior work, investigates user-level changes in perceptions as well as behaviors. When users make active choices, like deciding what timeline to view in their daily browsing, they are likely making those choices based on their beliefs about those feeds, rather than any hard evidence about their contents. The reality, according to our findings, that the \Algo timeline actually measures favorably against the \Chrono one in terms of information quality is likely unknown to most users. In particular, skeptical or mistrustful users basing their choices on their own perceptions (rather than our data) may therefore be making choices that inadvertently expose them to lower-quality content. Understanding how real users react both attitudinally and behaviorally to the algorithmic environment is crucial for improving overall outcomes; we advocate for further research to explore such algorithmic effects.

Second, while we saw swift and sizable changes in user behavior and information supplies, we also observed tremendous stability of user perceptions self-reported in surveys. This suggests that human attitude is more resilient to \Algo interventions than human behavior. Admittedly, it is quite possible that a longer-duration intervention is needed to move the needle, and that our three-week experiment is simply not long enough to make a significant change on user perceptions. However, an alternative and more pessimistic interpretation might be that people's perceptions of social media platforms, of political news, and of partisan others have been disconnected from objective information realities, instead becoming entrenched due to socio-psychological factors~\cite{fiske_social_1993}, with the result being that external interventions will have a difficult time changing people's minds. While this conclusion may soothe the concern surrounding digital news environments and their impact on democratic outcomes --- changes in information ecosystems can rapidly influence users' online behaviors but cannot significantly change their belief systems within a short period of time --- it begets another concern: that if user perceptions are so hard to change, the solution to the series of thorny social problems in the United States will require a much bigger systematic revamp than mere algorithmic changes of the variety tested in empirical studies like ours.

Third, and finally, the notable decrease in users' overall satisfaction in both experimental conditions compared to the observation week indicates users' distaste towards our timeline-enforcement intervention. Taking observational results into account, people exhibited a high level of agency in switching between the algorithmic and chronological timelines in their everyday, pre-intervention use of \Twitter, and they became disappointed when losing that agency. This finding offers insights for social media researchers and designers, highlighting that users should not be viewed as passive subjects merely influenced by the system and algorithms. Instead, they do possess autonomy in their interface usage and react sensitively to technical changes. Overall, this underscores the importance of the user-centric aspect sociotechnical audits for comprehending the social consequences of algorithmic systems. 


\subsection{Limitations}
We wish to recognize several limitations of this research. First, the sample we obtained would ideally be larger and more representative of \Twitter's user base. However, the former proved very challenging and expensive to access, and the latter may be impossible to achieve given the lack of solid statistics from the company. A related concern is that the crowdworkers we recruited from Prolific were not nationally representative~\cite{litman_reply_2021}. Driven by the research purpose of this study, we instead prioritized the creation of an ideologically balanced sample in our pre-screenings. While we tried to overcome these limitations with novel experimental designs and statistical inferences based on within-subject comparisons, the representativeness of the sample (or lack thereof) remains a point for improvement.

Second, this study did not probe into content-level analysis but only focused on the source level. We relied on Ad Fontes media bias scores, assigning source-based values to news URLs. Despite the high correspondence of available domain-level lists \cite{lin_high_2023}, this study would benefit from a robustness check with other lists of media bias labels. Our analysis also excluded users' exposure to other forms of political content --- for example if a news source was not listed in the Ad Fontes database, or if a tweet contained political discussions without any external news link. This is a common limitation in related work~\cite{huszar_algorithmic_2022,zhou_puzzle_2023}, one that will require additional methods and further research to resolve.

Third, although we did not inform users about the precise goal of the study, we cannot guarantee whether intervention effects were undermined by users' awareness, by the impact of payments on users' organic browsing behavior, or by carryover effects across weeks. This limitation is inherent to experimental designs with obtrusive interventions; although we believe user behavior is largely organic due to the length of the experiment, more research is needed to address the exact risk of such interventions on users' compliance and internal validity of this project.

Fourth, this study is limited in time and space: to only three weeks (perhaps not long enough to capture significant changes in user perceptions), and to desktop use of \Twitter. As a result, it does not account for mobile browsing data or changes over longer periods of time, which limits the generalization scope of our results~\cite{yang_exposure_2020}. Future work may involve overcoming the technical difficulties of collecting browsing data and enacting interventions on mobile devices to study the effects on mobile \Twitter users.









%% file: sections/7-conclusion.tex
\section{Conclusion}

Building on prior work from a range of disciplines on the state of the social media ecosystem with regards to news, we conduct a sociotechnical audit of news content on \Twitter. Recruiting \Twitter users (N=243) from across the political spectrum, we first passively observe user behavior and news exposure for one week on the platform, and subsequently conduct an intervention, enforcing that users spend another week restricted to the \Algo ``For You'' timeline and a third week restricted to the \Chrono ``Following'' timeline (randomizing the order of interventions). Analyzing our results, we find that (1) while the \Algo timeline decreases user exposure to news, it also has a moderating effect and users see less extreme, less like-minded and slightly more reliable news sources; and (2) despite our intervention having significant effects on users' behaviors, and users reporting sensitivity to the bias of content that they see, their broader attitudes and beliefs remained largely stable. 

Our first major take-away from this work was the surprise (especially in the era of Elon Musk) that the algorithmic timeline on \Twitter did not prioritize engagement at the expense of news quality; instead, engagement and quality were both higher (though the amount of news did decrease for users in this condition). While many have decried the algorithmic amplification of political content or hate speech, we found little evidence supporting these claims. While algorithms decrease the quantity of news available, some may simultaneously increase its quality. Our second major conclusion is that, while our algorithm audit shows that the system may perform well technically, our user-centered audit showed users remained stubborn in their beliefs about the system. This suggests a need to re-calibrate technical performance and user perceptions so they match more closely, beginning with centering and studying users.

%% file: sections/8-appendix.tex
\section{Appendix}
\subsection{Method: Survey Questions}
\label{appendix:survey}
\begin{enumerate}
    \item We'd like you to roughly estimate the ideological bias of the news on Twitter (X) on a 100-point scale for the past week. 0 means extremely liberal, 50 means ideologically neutral and 100 means extremely conservative.
\begin{enumerate}
\item "Recall the news on your Twitter (X) timeline in the past week:"
range 0 -100

\item All news available on Twitter (X) (including for you and other users) in the past week:
range 0 -100

\item News on a typical conservative user's Twitter (X) timeline in the past week:
range 0 -100

\item News on a typical liberal user's Twitter (X) timeline in the past week:
range 0 -100
\end{enumerate}
\item Thinking back on the last week, how satisfied are you with the user experience on Twitter (X) overall?  

5-point Likert scale (“extremely unsatisfied” to “extremely satisfied”).

\item How satisfied are you with the news environment on Twitter (X) in the last week? 

5-point Likert scale (“extremely unsatisfied” to “extremely satisfied”). 

\item This is an attention check. Please select 'Neither agree nor disagree' to indicate you are paying attention to survey questions. 

5-point Likert scale ("strongly disagree" to "strongly agree")

\item Please estimate the credibility of the news information you see on Twitter (X):  

5-point Likert scale ("extremely low credibility" to "extremely high credibility")

\item Please indicate the extent to which you trust the news information you see on Twitter (X):  

5-point Likert scale ("extremely distrust" to "extremely trust")

\item Do you think the Twitter (X) algorithm is generally neutral, or primarily supports posts from liberals or conservatives?  

5-point Likert scale ("primarily supports liberals' posts" to "primarily supports conservatives' posts")

\item To what extent do you agree with the following statement: There should be more government regulation of social media companies like Twitter (X)  

5-point Likert scale ("strongly disagree" to "strongly agree")

\item To what extent are you aware that algorithms are used to recommend content to the user on social media?
5-point Likert scale ("not at all aware" to "completely aware") 

\item To what extent are you aware that algorithms are used to prioritize certain content above others on social media?  

5-point Likert scale ("not at all aware" to "completely aware") 

\item To what extent are you aware that algorithms are used to tailor certain content to the user on social media?  

5-point Likert scale ("not at all aware" to "completely aware") 

\item To what extent are you aware that algorithms are used to show someone else different news than you get to see on social media?  

5-point Likert scale ("not at all aware" to "completely aware")  

\item Most news online is distorted by social media algorithms.  

5-point Likert scale ("strongly disagree" to "strongly agree")
\item News delivered through social media algorithms is harmful to democracy.  

5-point Likert scale ("strongly disagree" to "strongly agree")

\item Social media algorithms deliver news that tends to be slanted against my views.  

5-point Likert scale ("strongly disagree" to "strongly agree")
\end{enumerate}
\newpage

\subsection{Summary Statistics}
\label{appendix:survey-summary-statistics}
Perceived echo chamber measures are on a scale from 0 (extremely liberal) to 100 (extremely conservative). All other measures are on a 5-point likert scale (for details refer to Appendix~\ref{appendix:survey}).  We merged survey items (9) to (12) from Appendix~\ref{appendix:survey} as a multi-item measure of algorithmic media content awareness~\cite{zarouali_is_2021} (Cronbach's $\alpha > 0.9$ across all surveys).
\label{appendix:survey-summary-statistics}
\begin{table}[h!]
    \centering
  \caption{Averages and standard deviations for survey measures for Survey 2 (after Observation).}
  \label{tab:survey2-summary-statistics}
    \begin{tabular}{lllllll}
    \toprule
                                  & \multicolumn{2}{l}{Democrat} & \multicolumn{2}{l}{Independent} & \multicolumn{2}{l}{Republican} \\
                                  \midrule
                                  & Mean & SD & Mean & SD & Mean & SD \\
                                  \midrule
    Perceived Echo Chambers Own Timeline         & 36.83 & 24.82 & 45.68 & 27.91 & 63.95 & 20.77 \\
    Perceived Echo Chambers Overall   & 53.20 & 20.19 & 53.51 & 18.39 & 50.91 & 17.82 \\
    Perceived Echo Chambers Rep Timeline   & 82.00 & 20.37 & 74.06 & 17.21 & 69.17 & 18.48 \\
    Perceived Echo Chambers Dem Timeline   & 25.77 & 19.24 & 31.65 & 22.34 & 27.86 & 23.67 \\
    \midrule
    Satisfaction with User Experience   & 3.13 & 1.15 & 3.31 & 0.80 & 4.04 & 0.92 \\
    Satisfaction with News   & 2.79 & 1.10 & 3.06 & 0.92 & 3.82 & 0.91 \\
    Perceived News Credibility   & 2.75 & 0.95 & 3.03 & 0.86 & 3.46 & 0.91 \\
    Trust in News   & 2.85 & 1.04 & 2.96 & 0.98 & 3.53 & 0.86 \\
    Perceived Neutrality of Platform   & 3.55 & 1.03 & 3.51 & 0.73 & 2.97 & 0.69 \\
    Support for Regulation   & 3.10 & 1.27 & 2.27 & 1.30 & 1.93 & 1.26 \\
    Algorithmic Media Content Awareness   & 4.23 & 0.86 & 3.85 & 1.23 & 4.08 & 0.91 \\
    Algorithms Distort News   & 3.96 & 0.86 & 3.58 & 0.98 & 3.64 & 1.00 \\
    Algorithm-delivered News Harm Democracy   & 3.23 & 1.07 & 2.82 & 1.07 & 2.60 & 1.11 \\
    Algorithm-delivered News is Slanted Against My Views   & 2.79 & 1.06 & 3.03 & 0.86 & 3.02 & 0.91 \\
    \bottomrule
    \end{tabular}
\end{table}

\begin{table}[h!]
    \centering
  \caption{Averages and standard deviations for survey measures for Survey 3 (after Intervention 1).}
  \label{tab:survey3-summary-statistics}
\begin{tabular}{lllllll}
\toprule
                              & \multicolumn{2}{l}{Democrat} & \multicolumn{2}{l}{Independent} & \multicolumn{2}{l}{Republican} \\
                              \midrule
                              & Mean & SD & Mean & SD & Mean & SD \\
                              \midrule
Perceived Echo Chambers Own Timeline         & 39.76 & 22.29 & 47.06 & 24.57 & 65.84 & 19.23 \\
Perceived Echo Chambers Overall   & 55.96 & 18.23 & 49.48 & 23.47 & 49.33 & 18.25 \\
Perceived Echo Chambers Rep Timeline   & 81.11 & 20.86 & 72.37 & 19.49 & 73.48 & 14.56 \\
Perceived Echo Chambers Dem Timeline   & 28.57 & 19.46 & 36.37 & 27.20 & 29.26 & 24.28 \\
\midrule
Satisfaction with User Experience   & 2.91 & 1.19 & 3.17 & 0.88 & 3.93 & 1.07 \\
Satisfaction with News   & 2.82 & 1.12 & 3.13 & 0.95 & 3.73 & 1.05 \\
Perceived News Credibility   & 2.74 & 0.99 & 3.06 & 0.92 & 3.44 & 1.01 \\
Trust in News   & 2.75 & 1.05 & 2.89 & 1.01 & 3.42 & 0.91 \\
Perceived Neutrality of Platform   & 3.60 & 1.01 & 3.48 & 0.82 & 2.93 & 0.57 \\
Support for Regulation   & 3.28 & 1.21 & 2.44 & 1.52 & 1.88 & 1.15 \\
\bottomrule
\end{tabular}
\end{table}

\begin{table}[]
    \centering
  \caption{Averages and standard deviations for survey measures for Survey 4 (after Intervention 2).}
  \label{tab:survey4-summary-statistics}
\begin{tabular}{lllllll}
\toprule
                              & \multicolumn{2}{l}{Democrat} & \multicolumn{2}{l}{Independent} & \multicolumn{2}{l}{Republican} \\
                              \midrule
                              & Mean & SD & Mean & SD & Mean & SD \\
                              \midrule
Perceived Echo Chambers Own Timeline         & 43.00 & 24.43 & 50.60 & 22.21 & 64.02 & 15.77 \\
Perceived Echo Chambers Overall   & 54.27 & 20.16 & 57.12 & 16.25 & 50.28 & 17.92 \\
Perceived Echo Chambers Rep Timeline   & 81.08 & 17.49 & 68.60 & 22.26 & 68.88 & 20.28 \\
Perceived Echo Chambers Dem Timeline   & 34.88 & 21.64 & 31.96 & 23.91 & 33.77 & 27.23 \\
\midrule
Satisfaction with User Experience   & 2.93 & 1.21 & 3.16 & 0.98 & 4.13 & 0.86 \\
Satisfaction with News   & 2.75 & 1.16 & 3.04 & 0.84 & 3.73 & 1.00 \\
Perceived News Credibility   & 2.60 & 0.95 & 3.04 & 1.05 & 3.42 & 1.03 \\
Trust in News   & 2.70 & 1.12 & 3.04 & 1.01 & 3.55 & 0.96 \\
Perceived Neutrality of Platform   & 3.59 & 0.95 & 3.36 & 0.86 & 3.08 & 0.79 \\
Support for Regulation   & 3.15 & 1.33 & 2.32 & 1.40 & 1.88 & 1.19 \\
Algorithmic Media Content Awareness   & 4.23 & 0.88 & 3.90 & 1.06 & 4.04 & 0.99 \\
Perceptions of Algorithmic News Bias   & 3.43 & 1.13 & 3.01 & 1.12 & 3.10 & 1.13 \\
Algorithms Distort News   & 4.05 & 0.84 & 3.56 & 1.08 & 3.62 & 1.00 \\
Algorithm-delivered News Harm Democracy   & 3.32 & 1.09 & 2.52 & 1.08 & 2.53 & 1.19 \\
Algorithm-delivered News is Slanted Against My Views   & 2.91 & 1.12 & 2.96 & 0.97 & 3.13 & 0.89 \\
\bottomrule
\end{tabular}
\end{table}
\clearpage


\subsection{Participant Demographics}
\label{appendix:demographics}
\begin{figure}[ht]
  \centering
  \includegraphics[width=\linewidth]{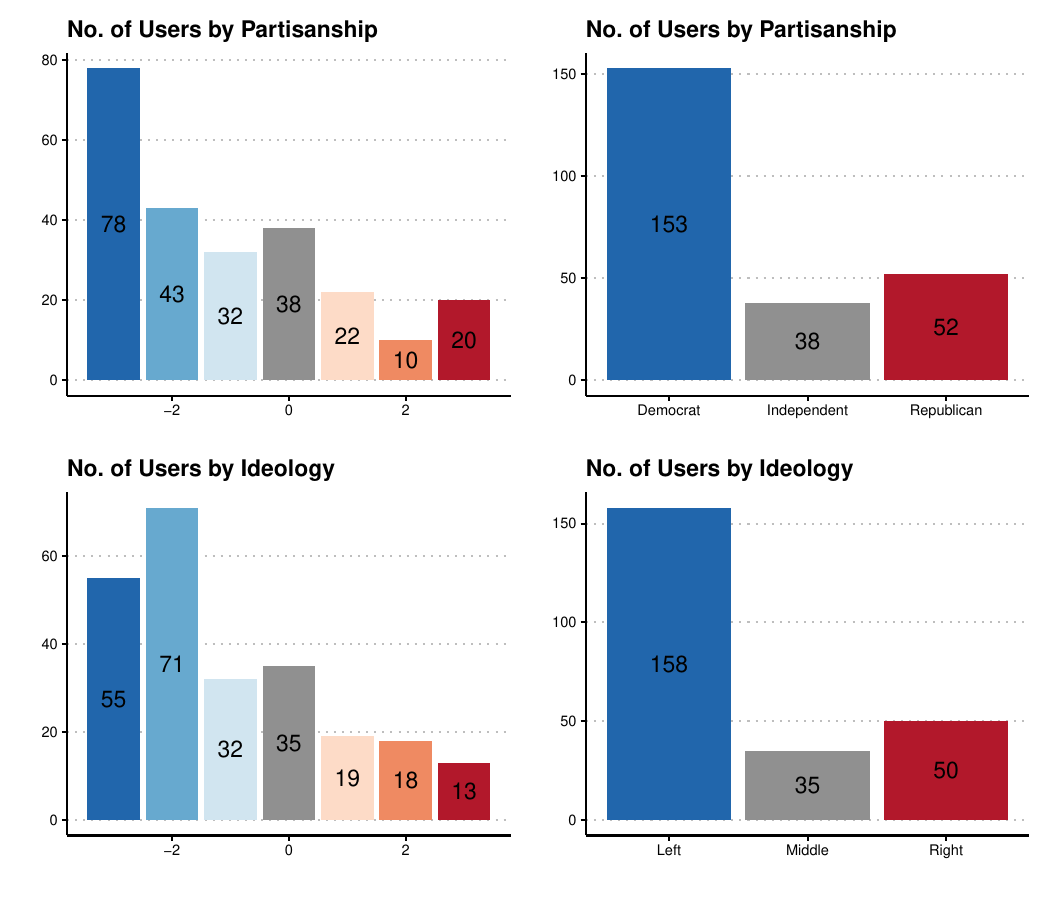}
  \caption{Summary of participants (N=243) by self-reported partisanship and ideology. Partisanship is reported on a 7-point scale from ('strong democrat' to 'strong republican'). Ideology is reported on a 7-point scale from ('extremely liberal' to 'extremely conservative').}
  \Description{Four bar charts breaking down participant numbers by self-reported partisanship and ideology. The two left-most charts show the distribution over 7-point categories. The two right-most charts show the distribution bucketed into 3-point categories.}
  \label{fig:userdescription1}
\end{figure}
\newpage
\begin{figure}[ht]
  \centering
  \includegraphics[width=0.75\linewidth]{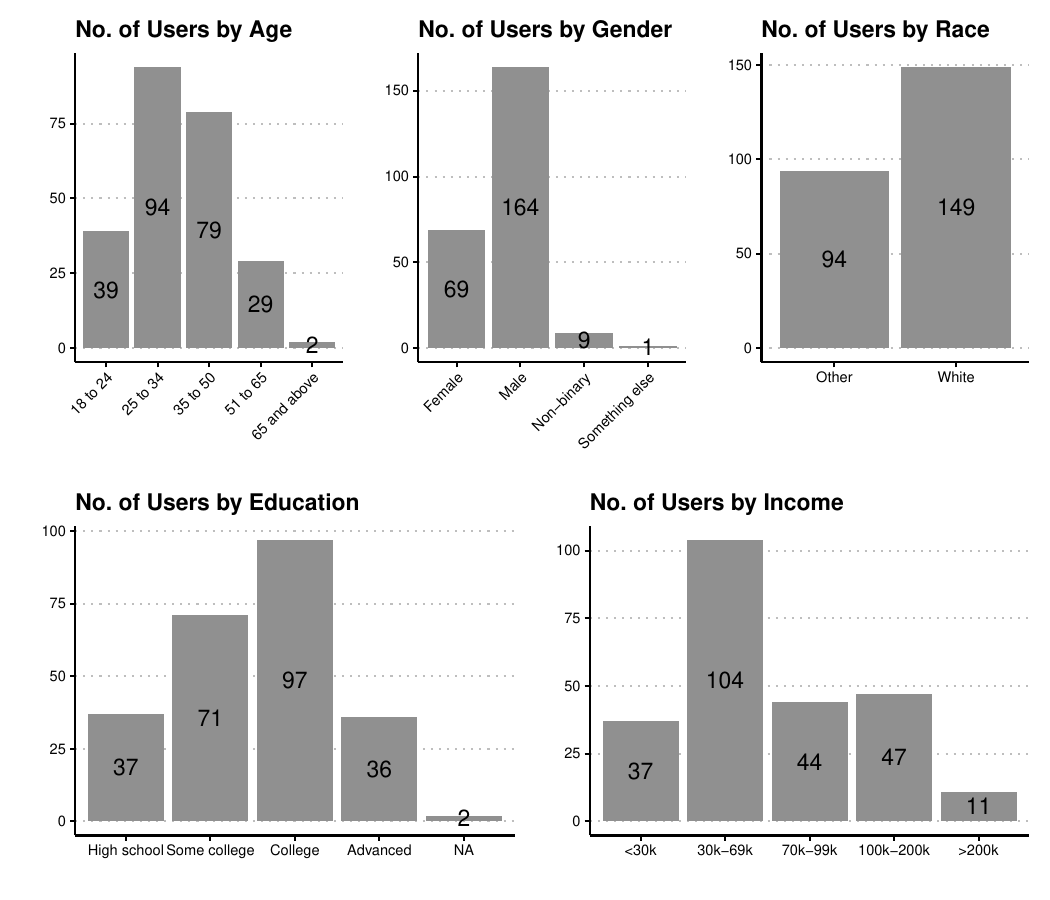}
  \caption{Summary of participant demographics (N=243) by age, gender, race, education level and income level.}
  \Description{Five bar charts showing participant demographics by age, gender, race, education level and income level.}
  \label{fig:userdescription2}
\end{figure}
